\documentclass[amssymb,aps,amsmath,floatfix,prfluids,superscriptaddress,longbibliography]{revtex4-2}
\usepackage[latin1]{inputenc}

\usepackage{mathptmx}
\usepackage[T1]{fontenc}
\usepackage{graphicx,overpic}
\usepackage{amsmath,amssymb,amsfonts, MnSymbol}
\usepackage{mathrsfs}
\usepackage{bm}
\usepackage{hyperref}
\usepackage{bbold}
\usepackage{xcolor}
\usepackage{tikz}
\usepackage{tabularx}

\newcommand{\algn}[1]{\begin{align} #1 \end{align}}
\newcommand{\sbeqs}[1]{\begin{subequations} #1 \end{subequations}}
\newcommand{\pmat}[1]{\begin{pmatrix} #1 \end{pmatrix}}

\newcommand{\tr}{\text{Tr}}
\newcommand{\eps}{\ensuremath{\varepsilon}}
\newcommand{\taup}{\ensuremath{\tau_\text{p}}}
\newcommand{\tauk}{\ensuremath{\tau_\text{K}}}
\newcommand{\tauc}{\ensuremath{\tau_\text{c}}}

\newcommand{\ve}[1]{\bm{#1}}
\newcommand{\ma}[1]{\ensuremath{\mathbb{#1}}}

\newcommand{\ee}{\ensuremath{\text{e}}}
\newcommand{\ed}{\ensuremath{\text{d}}}

\newcommand{\tdd}[1]{\ensuremath{\tfrac{\text{d}}{\text{d} #1}}}

\newcommand{\mc}[1]{\ensuremath{\mathcal{#1}}}
\newcommand{\mbb}[1]{\ensuremath{\mathbb{#1}}}
\newcommand{\ms}[1]{\ensuremath{\mathscr{#1}}}
\newcommand{\st}{\ensuremath{\text{St}}}

\newcommand{\eqnlab}[1]{\label{eq:#1}}
\newcommand{\seclab}[1]{\label{sec:#1}}
\newcommand{\figlab}[1]{\label{fig:#1}}
\newcommand{\eqnref}[1]{\eqref{eq:#1}}
\newcommand{\Eqnref}[1]{Eq.~\eqref{eq:#1}}
\newcommand{\Eqsref}[1]{Eqs.~\eqref{eq:#1}}

\newcommand{\Secref}[1]{Section~\ref{sec:#1}}
\newcommand{\figref}[1]{\ref{fig:#1}}
\newcommand{\Figref}[1]{Fig.~\ref{fig:#1}}

\begin{document}
\title{Caustic formation in a non-Gaussian model for turbulent aerosols}

\author{J. Meibohm}
\affiliation{Department of Mathematics, King's College London, London WC2R 2LS, United Kingdom}
\author{L. Sundberg}
\affiliation{Department of Physics, Gothenburg University, SE-41296 Gothenburg, Sweden}
\author{B. Mehlig}
\affiliation{Department of Physics, Gothenburg University, SE-41296 Gothenburg, Sweden}
\author{K. Gustavsson}
\affiliation{Department of Physics, Gothenburg University, SE-41296 Gothenburg, Sweden}

\begin{abstract}
Caustics in the dynamics of heavy particles in turbulence  accelerate particle collisions. The rate $\ms J$ at which these singularities form depends sensitively on the Stokes number $\st$,  the non-dimensional inertia parameter. Exact results for this sensitive dependence have been obtained using Gaussian statistical models for turbulent aerosols. However, direct numerical simulations of heavy particles in turbulence yield much larger caustic-formation rates than predicted
by the Gaussian theory. In order to understand possible mechanisms explaining this difference, we analyse  a non-Gaussian statistical model for caustic formation in the limit of small $\st$. We show that at small $\st$, $\ms J$  depends sensitively on the tails of the distribution of Lagrangian fluid-velocity gradients. This explains why different authors obtained different $\st$-dependencies of $\ms J$ in numerical-simulation studies. The most-likely gradient fluctuation that induces caustics at small $\st$, by contrast, is the same in the non-Gaussian and Gaussian models. Direct-numerical simulation results for particles in turbulence show that the optimal fluctuation is similar, but not identical, to that obtained by the model calculations.
\end{abstract}

\maketitle
\section{Introduction}
Particle inertia causes heavy particles suspended in an incompressible turbulent flow to cluster and form fractal spatial patterns~\cite{bec2003fractal}, because it allows the particles to detach from the flow. This happens at caustics, singular points of the particle dynamics, where the volume of the infinitesimal particle neighbourhood collapses to zero~\cite{crisanti1992lagrangian,falkovich2002acceleration,wilkinson2005caustics,gustavsson2016statistical,bec2024statistical}. Between caustics, the particle velocities are multi-valued, giving rise to anomalously large relative particle-velocities at small separations~\cite{falkovich2002acceleration,bec2010intermittency,salazar2012inertial,gustavsson2014relative}. This effect is also known as the ``sling effect'' \cite{falkovich2002acceleration,falkovich2007sling,bewley2013observation}. It causes spatial continuum descriptions of the inertial particle velocity field to fail. 
To some degree, the nonsmooth contribution from multivaluedness can be modeled using spatially uncorrelated fluctuations~\cite{fevrier2005partitioning}, or more accurate approaches~\cite{ducasse2009inertial,ijzermans2010segregation,papoutsakis2018modelling}.

Theoretical analysis of Gaussian statistical models resolving individual particle trajectories shows that, at small Stokes numbers, caustics form by an optimal fluctuation that involves
large fluid strain and zero fluid vorticity~\cite{meibohm2021paths}. In the plane
spanned by the invariants $Q = -\frac12 \tr \ma A^2$ and $R = -\det \ma A$ of the traceless matrix $\ma A$ of fluid-velocity gradients, the optimal fluctuation that induces a caustic
follows the right branch of the Vieillefosse line $\frac{27}4 R^2 + Q^3=0$~\cite{meibohm2023caustics}. However, turbulent fluid-velocity gradients are not Gaussian distributed, which raises the question to which extent non-Gaussian fluctuations change these results.
\begin{figure}[t]
\begin{overpic}[width =\textwidth]{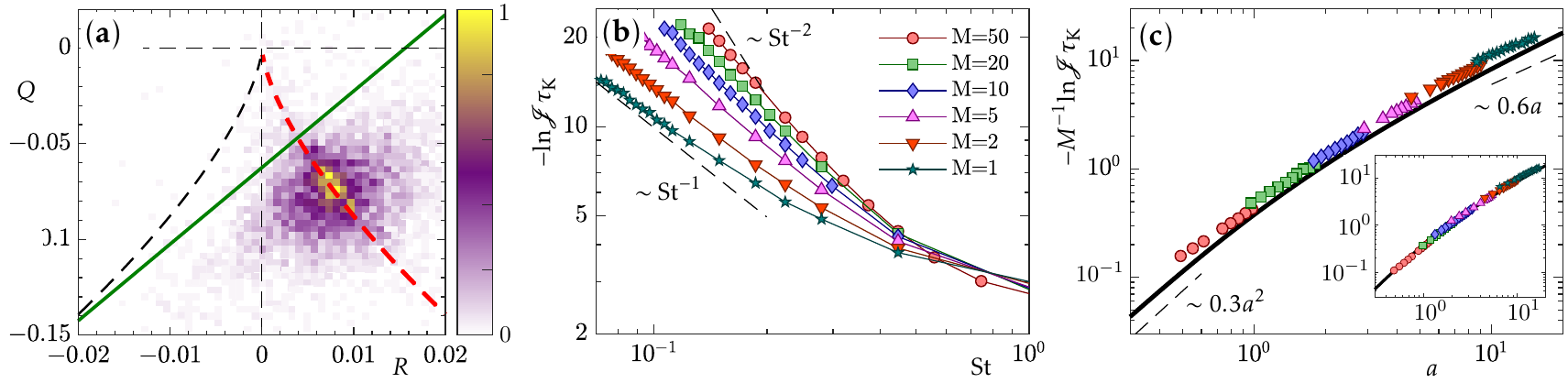}
\end{overpic}
        \caption{\label{fig:1} Caustic formation in a non-Gaussian model for a turbulent aerosol. ({\bf a}) Probability of $Q$ and $R$ at the onset of caustic formation (colour coded). Parameters: $\st=0.1$ and $M=1$. The thick dashed line
        is the right branch of the Vieillefosse  line (see text). The solid line is the threshold line for caustic formation, Eq.~(\ref{eq:threshold_line}).
      ({\bf b}) Rate of caustic formation for the non-Gaussian model for different values of $M$. Also shown are the limiting $\st$ scalings, dashed lines. ({\bf c}) Collapse of the small-$\st$ data from panel ({\bf b}) (markers) onto the scaling function $\ms{F}(a)$ given in \Eqnref{ratefunc} (thick black line), using the $\st\to 0$ limit of the scaling variable $a$ in \Eqnref{J2}. The inset shows the same, but using the finite Stokes correction for $a$ in \Eqnref{scalvar} with $\delta\lambda_{\rm th}$ from the theory in Table~\figref{scaling}.
       }
\end{figure}

To address this question, we formulate a statistical model that exhibits non-Gaussian fluid-velocity statistics and analyse caustic formation in the small-$\st$ limit~\cite{derevyanko2007lagrangian,meibohm2017relative,meibohm2019heavy,meibohm2021paths,meibohm2023caustics} using optimal-fluctuation theory. We find that
the optimal fluctuation agrees with that in the Gaussian models at small $\st$: caustics form along the right branch of the Vieillefosse line [Fig.~\ref{fig:1}({\bf a})].  The rate of caustic formation $\mathscr{J}\tauk$, by contrast, depends strongly on the shape of the non-Gaussian tails of the distribution of $\ma A$. At small Stokes numbers, its $\st$-dependence is determined by a non-universal action that is sensitive to the tails of the distribution of fluid-velocity gradients [Fig.~\ref{fig:1}({\bf b})]. As $\st\to 0$, the rate of caustic formation obeys an asymptotic law in the form
\begin{equation}
\label{eq:J2}
-\ln(\mathscr{J}\tauk)\sim M\ms{F}(a)\,\quad\text{with}\quad a=\frac{\sqrt{d(d+2)}}{4\st\sqrt{M}}  +\mc{O}(\st^{-1/3})\,,
\end{equation}
in $d$ spatial dimensions. Here, $\ms{F}$ is a scaling function that we determine explicitly and $M\geq1$ is a model parameter described in \Secref{model}. Figure~\figref{1}({\bf c}) shows the collapse of the data onto the scaling function $\ms{F}(a)$ for small enough $\st$. For $\st \sqrt{M}\gg 1$, the non-Gaussian model coincides with the Gaussian model and we find inversely quadratic scaling in $\st$, $-\ln(\mathscr{J}\tauk)\propto \st^{-2}$, in agreement with earlier models~\cite{falkovich2002acceleration,derevyanko2007lagrangian,gustavsson2013distribution}. For $\st \sqrt{M}\ll 1$, by contrast, the $\st$ scaling is inversely linear, $-\ln(\mathscr{J}\tauk)\propto \st^{-1}$. The different $\st$-scalings arise because the rate of caustic formation is determined by the tails of the probability distribution of the fluid-velocity gradients, which depend on the relative magnitudes of $M$ and $\st$. The incomplete collapse of the data in \Figref{1}({\bf c}) is a finite-$\st$ effect discussed in \Secref{datacoll}, as indicated by the correction term $\mc{O}(\st^{-1/3})$ in \Eqnref{J2}. Including the leading-order correction in $\st$ (\Secref{datacoll}) improves the collapse, see inset of \Figref{1}({\bf c}).

In contrast to the sensitive dependence of $\ms{J}$ on the parameters, the most likely history of fluid-velocity gradients that induces a caustic -- the optimal fluctuation -- is the same for all parameters $M$ of the non-Gaussian model, and identical to that of the Gaussian model~\cite{meibohm2021paths,meibohm2023caustics} for small $\st$. This indicates that the optimal fluctuation that induces a caustic is more robust than the rate of its occurrence, which is strongly model dependent. To emphasise this point, we compare our results with direct numerical simulations (DNS) of particles in turbulence and find qualitative agreement for the optimal fluctuation, providing further evidence for its robustness.

Our results explain why studies aimed at comparing the rate of caustic formation $\ms{J}$ from DNS with that from idealised (and often Gaussian) models fail: For small $\st$, $\ms{J}$ depends of the tails of the distribution of fluid-velocity gradients which is non-Gaussian and unknown in general.
\section{Caustic formation}\label{sec:caustics}
In the Stokes approximation, the equation of motion for a small, yet heavy spherical particle reads~\cite{bec2024statistical}
\algn{\label{eq:st}
	\dot{\ve x} = \ve v\,,\quad\dot{\ve v} = {\taup^{-1}}(\ve u(\ve x,t)-\ve v)\,.
}
Here $\ve x$ and $\ve v$ are particle position and velocity, $\ve u(\ve x,t)$ is the turbulent velocity at the particle position at time $t$, and dots denote time derivatives.
Gravitational settling is neglected (the effect of settling on caustic formation is discussed in Refs.~\cite{bec2014gravity,gustavsson2014clustering}).
The Stokes approximation applies for small enough particles that are much heavier than the fluid. In this limit, particle inertia can nevertheless be important, even when the particles are small. Caustics occur when particle-velocity field folds over configuration space, so that the particle-velocity gradients $\partial v_i/\partial x_j$  tend to $-\infty$ \cite{gustavsson2016statistical,meibohm2021paths,meibohm2023caustics}.
To identify caustics, one therefore follows the particle-velocity gradients and determines when they diverge. To this end, one integrates~\cite{falkovich2002acceleration}
\algn{\label{eq:Z}
	\taup\dot {\ma Z} = \ma A(\ve x, t)-\ma Z-\ma Z^2\,,
}
  alongside \Eqnref{st}.
The matrices $\ma A$ and $\ma Z$ contain the fluid-velocity gradients $A_{ij} = {\taup}\partial u_i/\partial x_j$ and particle-velocity gradients $Z_{ij} = {\taup}\partial v_i/\partial x_j$ at the particle position, non-dimensionalised with the particle-response time $\taup$.
Equations~\eqnref{st} and \eqnref{Z}  are used to identify caustic locations. Approaches that aim to resolve the particle number density close to a caustic, by contrast, require also higher-order derivatives of the particle velocity \cite{papoutsakis2018modelling,papoutsakis2020a}.

We consider caustic formation at weak particle inertia, $\st\ll1$. In this case, the dynamics of $\ma Z$ occurs on a time scale of order $\taup$, much shorter than $\tauk$, the time scale of changes of $\ma A$. In this ``persistent limit''~\cite{derevyanko2007lagrangian,meibohm2017relative,meibohm2019heavy,meibohm2021paths,meibohm2023caustics}, changes of $\ma A$ can be treated as approximately constant. In this approximation, caustics occur whenever \Eqnref{Z} has no stable fixed points, because then $\mbb{Z}$ escapes to negative infinity in finite time. In three spatial dimensions, the fixed points of \Eqnref{Z} vanish whenever the fluid-velocity gradients $\mbb{A}$ exceed a threshold in the $Q$-$R$ plane, parameterised by the line~\cite{meibohm2023caustics}
 \begin{equation}
 \label{eq:threshold_line}
 Q = 4R-\tfrac{1}{16} \quad\mbox{for} \quad R \geq -\tfrac{1}{32}\,.
 \end{equation}
In the chosen non-dimensionalisation, the elements of $\ma A$ are typically of the order of $\st$, and thus small when $\st\ll 1$. Since the threshold~\eqnref{threshold_line} is of order unity, this means that rare large fluctuations of $\mbb{A}$ are needed to reach the threshold line.

An equivalent condition can be formulated in terms of the eigenvalues of $\mbb{A}$~\cite{meibohm2021paths,meibohm2023caustics,Bat23}. It requires that the most negative eigenvalue of $\mbb{A}$ be real and drop below a negative threshold that tends to $-1/4$ as $\st\to0$. In contrast to \Eqnref{threshold_line}, this condition is independent of the spatial dimension $d$.

Inertialess tracers move along the stream lines of the flow and sample configuration space homogeneously~\cite{falkovich2001particles}. As a consequence, the Lagrangian statistics of $\mbb{A}$ agrees with the Eulerian one. Inertial particles, by contrast, preferentially sample straining regions~\cite{maxey1987gravitational} and distribute inhomogeneously over configuration space. This leads to different magnitudes of the Lagrangian correlation functions of
of strain
$\ma S = (\ma A + \ma A^{\sf T})/2$ and vorticity $\ma O = (\ma A - \ma A^{\sf T})/2$ for finite $\st$~\cite{girimaji1990diffusion,brunk1997hydrodynamic,zaichik2003pair,vincenzi2007stretching},
\begin{subequations}\eqnlab{CSOeqs}
\algn{
	&\langle S_{ij}(t)S_{kl}(t')\rangle = \frac{C_S(\st) \st^2}{d(d+2)(d-1)}\left[d(\delta_{ik}\delta_{jl}+\delta_{il}\delta_{jk}) - 2\delta_{ij}\delta_{kl}\right]f_S(t-t')\,,\\
	&\langle O_{ij}(t)O_{kl}(t')\rangle = \frac{C_O(\st) \st^2}{d(d-1)}\left(\delta_{ik}\delta_{jl}-\delta_{il}\delta_{jk}\right)f_O(t-t')\,,\quad \langle O_{ij}S_{kl}\rangle = 0\,.
}
\end{subequations}
Here, $f_S(x)$ and $f_O(x)$ are normalised correlation functions with $f_S(0)=f_O(0)=1$ and $C_S(\st)$ and $C_O(\st)$ quantify the relative magnitude of strain and vorticity along inertial particle trajectories. Preferential concentration~\cite{squires1991preferential,eaton1994preferential} leads to $C_S(\st)>C_O(\st)$, i.e., strain dominates at finite $\st$~\cite{maxey1987gravitational}. However,  preferential concentration is absent for very small $\st$, so that $C_S(\st)=C_O(\st)=1/2$ for $\st\to0$.

 For Gaussian-distributed fluid-velocity gradients, Eqs.~(\ref{eq:CSOeqs}) imply that the steady-state probability $P_s(\mbb{A})$ for the occurrence of a realisation $\mbb{A}$ of fluid-velocity gradients is given by
 \begin{align}
 \label{eq:PAG}
 -\ln P_s(\ma A ) \sim
  & \frac{d-1}{4\st^2}\left(\frac{d+2}{C_S}\tr \mbb{S}^{\!\!\sf T}\!\!\mbb{S} + \frac{d}{C_O}\tr\mbb{O}^{\!\!\sf T}\!\!\mbb{O}\right)\,,
\end{align}
where we dropped the normalisation factor because it is subleading in the limit of small $\st$. To leading order in $\st$, the logarithmic probability for $\ma A$ to reach the threshold $\ma A_{\rm th}$  determines the rate of caustic formation, $-\ln (\ms J \tauk) \sim -\ln P(\ma A_{\rm th})$. For Gaussian fluid-velocity gradients \eqnref{PAG}, this leads to the inverse quadratic scaling in $\st$ found in Refs.~\cite{meibohm2021paths,meibohm2023caustics}, $-\ln (\ms J \tauk) \propto \st^{-2}$ as $\st\to0$.
\section{Non-Gaussian model}\label{sec:model}
Motivated by the fact that turbulent fluid-velocity gradients have non-Gaussian tails, we explore the effects of such tails on caustic formation. To this end, we formulate a non-Gaussian model for the turbulent fluid-velocity gradients $\ma A$ driving  Eq.~(\ref{eq:Z}). It is based on an ensemble of $M$ identical, time-independent random velocity fields $\ve u_m(\ve x)$, all with identical smooth spatial correlation functions. We superpose these $\ve u_m(\ve x)$ with random, time-dependent coefficients $c_m(t)$ to obtain a non-Gaussian model for turbulent fluctuations,
\begin{equation}\label{eq:3Dmodel}
\ve u(\ve x,t) = \frac1{\sqrt{M}}\sum_{m=1}^M c_m(t) \ve u_m(\ve x)\,,\qquad \ma{A}(\ve x,t) = \taup\ve \nabla \ve u(\ve x,t)\,.
\end{equation}
Here $c_m(t)$ are independent, identical Gaussian processes with zero mean and covariance
\algn{\eqnlab{ccorr}
        \langle c_m(t) c_n(t') \rangle = \delta_{mn} f(t-t')\,,
}
where $f(0)\!=\!1$ and with the correlation time $\tauc$ of $\ve u(\ve x,t)$ defined as ${\tauc} = \int_0^\infty\!\!\ed t f(t)$. For $\sqrt{M}\gg |A_{ij}|$, the central-limit theorem implies that $\mbb{A}$ is Gaussian distributed and $P_s(\mbb{A})$ is given by \Eqnref{PAG}, so that the results of \cite{meibohm2021paths,meibohm2023caustics} are recovered. For large fluid-velocity gradients and finite $M$, however, the central-limit theorem does not apply. In this case, the distribution of $\mbb{A}$ has non-Gaussian tails, whose form we discuss in the next section.
\section{Optimal-fluctuation theory}\label{sec:OFT}
For finite $M$ and large fluid-velocity gradients, the central limit theorem cannot be not applied to \Eqnref{3Dmodel}. In this case, we find by a saddle-point analysis of the non-Gaussian model
 \begin{align}
 \label{eq:PAnG}
 -\ln P_s(\ma A ) \sim\sqrt{\tfrac{M(d-1)}{2\st^2}\Big(\tfrac{d+2}{C_S}\tr\mbb{S}^{\!\!\sf T}\!\!\mbb{S}+ \tfrac{d}{C_O}\tr\mbb{O}^{\!\!\sf T}\!\!\mbb{O}\Big)}\,.
\end{align}
The derivation of \Eqnref{PAnG} is summarised in  Appendix \ref{app:A}. We see that the argument of the square root in \Eqnref{PAnG} is just $2M$ times the Gaussian action on the right-hand side of \Eqnref{PAG}, which simplifies the analysis that follows.

As mentioned earlier, at small $\st\ll1$, caustics form whenever the most negative eigenvalue of $\mbb{A}$ is real and drops below a negative threshold that tends to $-1/4$ for $\st\to0$~\cite{meibohm2021paths,meibohm2023caustics,Bat23}. In the persistent limit, the non-dimensional rate of caustic formation $\ms{J}\tauk$ is equal to the probability for this event, $P_s(\mbb{A}_\text{th})$, to leading order in $\st$. In order to compute this probability, we first express the right-hand side of \Eqsref{PAG} and \eqnref{PAnG} in terms of the eigenvalues $\ve{\lambda} = (\lambda_1,\ldots,\lambda_{d})^{\sf T}$ of $\mbb{S}$ and in terms of the non-zero singular values $\ve{\mu} = (\mu_1,\ldots,\mu_{\lfloor d/2\rfloor})^{\sf T}$ of $\mbb{O}$,
\algn{\eqnlab{actionlam}
	-\ln P_s(\ma A ) \sim -\ln P_s(\ve \lambda,\ve \mu )\sim\ms H\Big[\tfrac{d-1}{4\st^2}\Big(\tfrac{d+2}{C_S}|\ve \lambda|^2{+ \tfrac{2d}{C_O}|\ve \mu|^2}\Big)\Big]\,,}
where $\ms{H}(x)$ is equal to $x$ for $|A_{ij}|\ll\sqrt{M}$ and equal to $\sqrt{2M x}$ for $|A_{ij}|\gg\sqrt{M}$.
Transforming the joint probability density of $\mbb{A}$ into the probability of $\ve \lambda$ and $\ve \mu$ in this way involves a Jacobian determinant. This so-called Vandermode determinant, however, is independent of $\st$ and does therefore not contribute at this order. We marginalise the distribution~\eqnref{actionlam} over $\lambda_1,\ldots,\lambda_{d-1}$ and $\mu_1,\ldots,\mu_{\lfloor d/2 \rfloor}$ using the contraction principle~\cite{touchette2009large}
and compute
\algn{\eqnlab{margaction}
	-\ln P_s(\lambda_d )\sim\min_{\substack{\lambda_1,\ldots,\lambda_{d-1}, \ve \mu\\ \sum_{n=1}^d\lambda_n = 0}}\ms H\Big[
  & \tfrac{d-1}{4\st^2}\Big(\tfrac{d+2}{C_S}|\ve \lambda|^2{+ \tfrac{2d}{C_O}|\ve \mu|^2}\Big)\Big]\,,
}
by constrained minimisation.
The constraint $\sum_{n=1}^d\lambda_n = 0$ arises because $\mbb{A}$ must be traceless in incompressible flow. We enforce the constraint by adding a Lagrange multiplier.  Evaluating the stationary point of the resulting Lagrangian gives $\lambda_k^* = -\lambda_d/(d-1)$ and $\mu_j^*=0$,
independently of $\ms{H}$. This implies that the optimal gradient configuration that leads to a caustic has vanishing vorticity, $\ma O^*\equiv 0$, as shown in Refs.~\cite{meibohm2021paths,meibohm2023caustics}, while one of the eigenvalues of $\mbb{S}^*$, $\lambda_d$ here, becomes large and negative. All other eigenvalues $\lambda_k$ are most likely equal and positive, and of $1/(d-1)$ times the magnitude of $\lambda_d$, in keeping with the incompressibility constraint. Hence after diagonalisation and ordering of the eigenvalues, the most likely threshold configuration to induce a caustic has the form
\algn{\eqnlab{Ath}
	\mbb{A}^*_\text{th} = -\lambda_\text{th} \mbb{e}\,,
}
where $\lambda_\text{th}$ is the threshold value for the most-negative eigenvalue of $\mbb{A}^*_\text{th}$ and $\mbb{e}$ is the diagonal matrix
\algn{\eqnlab{elementd}
	\mbb{e} = \frac1{d-1}\pmat{-1&&&&\\&-1&&&\\&&\ldots&&\\&&&-1&\\&&&&d-1}\,.
}
For $d=3$, the fact that all subleading eigenvalues $\lambda_k^*$ are identical, $\lambda_k^* =\lambda_\text{th}/(d-1)$, implies that caustics are most likely induced by fluid-velocity gradients that lie on the Vieillefosse line, not only in the Gaussian~\cite{meibohm2023caustics} but also in the non-Gaussian model. Figure~\ref{fig:1}{(\bf a}) shows the excursion of fluid-velocity gradients in the $Q$-$R$ plane at a time of order $\tauk$ before caustic formation, obtained from simulations of the non-Gaussian model with $M=1$. We see that the fluid-velocity gradients prior to caustic formation lie beyond the threshold (solid line) and close to the Vieillefosse line (thick dashed line), as predicted by our analysis. Note that most gradients in \Figref{1}({\bf a}) lie considerably below the threshold line. The reason is that at non-zero $\st$ [$=0.1$ in \Figref{1}{(\bf a})], the threshold $\lambda_\text{th}$ is larger than for $\st\to0$, because the strict separation of time scales breaks down and caustic formation takes a finite time for $\st>0$. This means that the smallest eigenvalue of the matrix of fluid-velocity gradients must fall below $-1/4$ for a finite time in order to enable caustic formation~\cite{meibohm2021paths,meibohm2023caustics,Bat23}. Consequently, at finite $\st$, the threshold value $\lambda_\text{th}$ in \Eqnref{Ath} must be adjusted, i.e., $\lambda_\text{th} = 1/4+\delta\lambda_\text{th}$, where $\delta\lambda_\text{th}$ is a correction that vanishes as $\st\to0$. From the numerical solution of a model system, we show in \Secref{1D} that $\delta\lambda_\text{th}$ scales as $\delta\lambda_\text{th} \sim\st^{2/3}$ for $\st\ll1$.
\section{Shape of optimal fluctuation}\seclab{1D}
We now consider the shape of the optimal fluctuation $\mbb{A}^*(t)$ of fluid-velocity gradients as a function of time that causes the most negative eigenvalue to drop below $-\lambda_\text{th}$ at time $t_\text{th}$. To this end, we note that the matrix $\mbb{e}$ in \Eqnref{Ath} can be chosen as an element of an orthonormal basis of traceless matrices with respect to the inner product $\langle \mbb{X},\mbb{Y}\rangle = d^{-1}(d-1)\tr(\mbb{X}^{\sf T}\mbb{Y})$ defined by the trace $\tr$ of the product of two matrices $\mbb{X}$ and $\mbb{Y}$. The normalisation is chosen here such that $\langle \mbb{e},\mbb{e}\rangle = 1$. In such a basis, the most likely way to reach the threshold $\mbb{A}^*_\text{th}$ is by a large excursion of the amplitude $A^*(t)$ associated with the basis element $\mbb{e}$,
\algn{\eqnlab{Aonedim}
	\mbb{A}^*(t) = A^*(t)\mbb{e}\,,
}
while the amplitudes of all other other basis elements remain small. In this way, the amplitude $A^*(t)$ alone determines the shape of the optimal fluctuation $\mathbb{A}^*(t)$ for small $\st$, reflecting the effecitively one-dimensional nature of caustic formation at small $\st$~\cite{derevyanko2007lagrangian,meibohm2021paths,meibohm2023caustics,Bat23}. The eigenvalue threshold $-\lambda_\text{th}$ in \Eqnref{Ath} is then reached when $A^*(t)$ reaches $A^*(t_\text{th}) = -\lambda_\text{th}$ at time $t_\text{th}$.

\begin{figure}
\begin{overpic}[width =\textwidth]{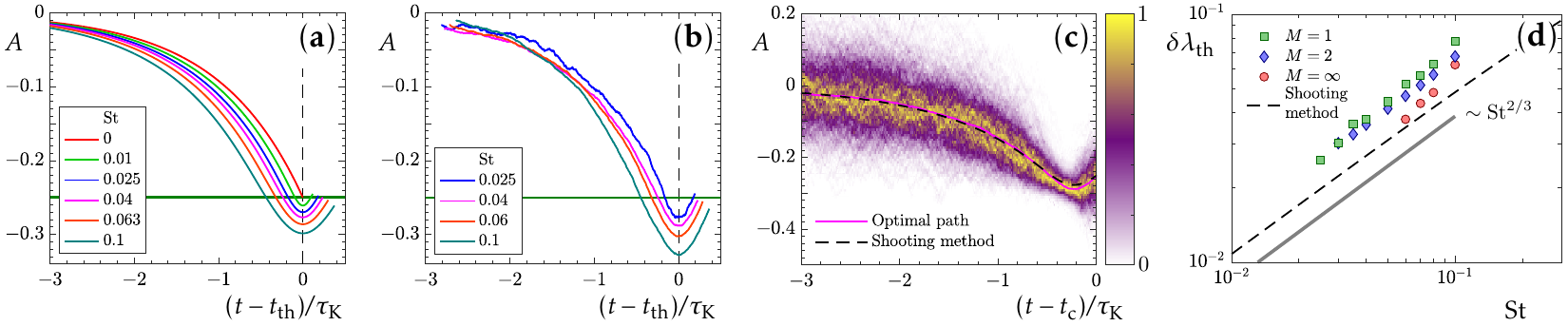}
\end{overpic}  %
	\caption{\label{fig:2} One-dimensional model for caustic formation.  ({\bf a}) Optimal fluctuations of fluid-velocity gradients, $A^*(t)$, obtained from the shooting method for the Gaussian case (see text) for different $\st$ vs. $t-t_\text{th}$. The thick horizontal line shows the threshold $-\lambda_{\rm th}=-1/4$. ({\bf b}) Same as ({\bf a}) but for simulations of the non-Gaussian model with $M=1$. The lines show the mean over realisations at each time step. ({\bf c}) Optimal fluctuation as a function of $t-t_{\rm c}$ using $M=1$ and $\st = 0.04$, where $t_{\rm c}$ is the time at which the caustic forms. The color map shows the probability distribution of $A(t)$, normalised at each time step by the maximal probability. The solid line is the mean obtained in panel ({\bf b}) and the dashed line shows the result from the shooting method in panel ({\bf a}).
	({\bf d}) The depth $\delta\lambda_\text{th}$ of the optimal fluctuation from simulations of the one-dimensional models with $M = 1,2$ and $\infty$.
	Also shown are the results of the shooting method (see text), as a dashed line.}
\end{figure}

In the previous Section, we discussed that the breakdown of time-scale separation increases the threshold value $\lambda_\text{th}(\st)$ for finite $\st$. Using the one-dimensional parametrisation~\eqnref{Aonedim}, we now make this statement more precise by considering the combined dynamics of the optimal fluctuations of fluid-velocity gradients and particle-velocity gradients. To this end, we model the optimal fluctuation $A^*(t)$ as the optimal fluctuation of an Ornstein-Uhlenbeck process. For this particular Gaussian process, the joint optimal fluctuations of the fluid-velocity gradient $A^*(t)$ and the escaping part of the particle-velocity gradient $Z^*(t)$ obey a closed set of differential equations, see e.g.~\cite{gustavsson2013distribution} for a derivation. These equations read
\sbeqs{\eqnlab{optfluctode}
\begin{align}
	\tdd{t}Z^*(t) &= \frac{1}{\st}\left(A^*-Z^*-Z^{*2}\right)\,,   \qquad \tdd{t}A^*(t)= -A^* + 2 \frac{\sigma_S^2}{\kappa_S}p_A\,,\\
	\tdd{t}p_z(t) &= \frac{1}{\st}(1+ 2Z^*)p_z\,,   \qquad \tdd{t}p_A(t) = \frac{p_A}{\kappa_S} - \frac{1}{\st}p_z\,,
\end{align}
}
where $0\leq t\leq t_\text{c}$ and $p_z$ and $p_A$ are conjugate momenta that enable non-trivial optimal fluctuations $A^*$ and $Z^*$. The parameter
\algn{\eqnlab{sigS}
	\sigma_S^2 = \frac{2C_S(\st)}{d(d+2)}\,,
}
and the non-dimensionalised Lagrangian strain correlation time $\kappa_S=\tauk^{-1}\int_0^\infty\!\!\ed t f_S(t)$ reflect the properties of fluid-velocity gradients along Lagrangian trajectories~\cite{girimaji1990diffusion,vincenzi2007stretching} in dimensions $d>1$. In order to generate a caustic, $Z^*$ in \Eqsref{optfluctode} must escape from the trivial fixed point $(Z^*,A^*,p_z,p_A) = (0,0,0,0)$ at $t=0$ and arrive at $Z^*\to-\infty$ at time $t=t_\text{c}$, the time at which a caustic forms. The condition that the fluctuation be optimal leads to the additional boundary conditions $p_z(t_\text{c}) = p_A(t_\text{c}) = 0$.

To find the optimal fluctuation that satisfies~\Eqsref{optfluctode} with the prescribed boundary conditions, we employ a numerical shooting method. Taking initial conditions on the unstable subspace of the trivial fixed point, we ``shoot'' trajectories obeying \Eqsref{optfluctode}, to ``hit'' the target boundary condition at $t=t_\text{th}$. Since the trivial fixed point has two unstable directions, we can parametrise the initial conditions by a single angle $\theta$ and write  $(Z^*,A^*,p_z,p_A) = \eps(\cos{\theta} \ve e_1 + \sin{\theta} \ve e_2)$, with $\eps\ll1$ at $t=0$. Here, the vectors $\ve e_1$ and $\ve e_2$ furnish an orthonormal basis of the unstable subspace of the trivial fixed point. By shooting trajectories that obey \Eqsref{optfluctode} with initial conditions parametrised by $\theta$, we optimise $\theta$ to hit the target boundary conditions $Z^*(t_\text{c})\to-\infty$, and $p_z(t_\text{c}) = p_A(t_\text{c}) = 0$. The trajectory $A^*(t)$ obtained in this way gives the amplitude $A^*(t)$ of the optimal fluctuation $\mbb{A}^*(t)$ in \Eqnref{Aonedim} for small but finite $\st$ when the fluid-velocity gradients are Ornstein-Uhlenbeck processes.

In order to understand how the optimal fluctuation obtained from the shooting method compares with the optimal fluctuation of the non-Gaussian model, we first consider a simplified, one-dimensional model with $\sigma_S^2 = \kappa_S = 1$. Figure~\figref{2}({\bf a}) shows the resulting amplitude $A^*(t)$ for different $\st$ plotted vs. $t-t_\text{th}$. We observe that the optimal fluctuation approaches the shape of the correlation function $A^*(t)\to -\frac14 f_S(t) = -\frac14 \exp(-|t-t_{\rm th}|/\kappa_S)$ of the Ornstein-Uhlenbeck process with $\kappa_S=1$ as $\st\to0$ [\Figref{2}({\bf a})]. This has been shown to be the case for all Gaussian processes in this limit~\cite{meibohm2021paths,meibohm2023caustics}. In Appendix~\ref{app:B} we show that this also holds for the non-Gaussian process studied here, which indicates that the optimal fluctuation of the non-Gaussian model approaches the correlation function in a similar way.

To verify that the optimal fluctuations of the non-Gaussian model for small $\st$ follow the same trend as observed in \Figref{2}({\bf a}) for the Ornstein-Uhlenbeck process, we compare $A^*(t)$ obtained from the shooting method with results of numerical simulations of the non-Gaussian model. To simplify the simulations and to thus allow for smaller $\st$, we consider again the one-dimensional version of the non-Gaussian model, obtained from by \Eqsref{3Dmodel} and \eqnref{Z} with $d=1$. In the one-dimensional model, the spatial dependence of the fluid-velocity gradients is neglected and the $c_m(t)$ are taken as independent Ornstein-Uhlenbeck processes with $\sigma_S^2 = \kappa_S = 1$. To obtain the optimal fluctuation, we integrate \Eqsref{3Dmodel} and \eqnref{Z} until a caustic forms and backtrace the history of the fluid-velocity gradient $A(t)$ prior to caustic formation at $t=t_\text{c}$. From the resulting gradient trajectories we compute for $M=1$ the mean, shown in \Figref{2}({\bf b}) for different values of $\st$, and the trajectory density, shown in \Figref{2}({\bf c}) for one small value of $\st$.
We observe that the trajectory density in \Figref{2}({\bf c}) sharply focuses in a narrow region of high density. This is the optimal fluctuation, which is well approximated by the mean value of the fluid-gradient at each time step (red line).

The mean values in \Figref{2}({\bf b}) show a similar trend as the Ornstein-Uhlenbeck process in \Figref{2}({\bf a}), approaching the shape of the correlation function as $\st$ becomes smaller. Moreover, \Figref{2}({\bf b}) shows that for decreasing $\st$ the threshold time $t_\text{th}$ where $A^*(t)$ has its minimum approaches the caustic formation time $t_\text{c}$. This implies that after a caustic has been initiated at $t=t_\text{th}$, it takes shorter and shorter time for it to form at $t=t_\text{c}$ as $\st$ decreases. Thus, as $\st\to0$ caustics form instantaneously whenever the amplitude $A^*(t)$ reaches $-1/4$~\cite{meibohm2021paths,meibohm2023caustics}. Note also that the smaller $\st$, the larger the magnitude of $A^*(t)$ at $t=t_\text{c}$. Since the optimal fluctuation $A^*(t)$ is a pure strain in our model, this implies that caustic formation events at small $\st$ are strongly correlated with straining regions of the flow. This provides an explanation for increased particle collisions in straining regions observed in Refs.~\cite{perrin2014preferred,picardo2019flow,Lee23}.

The mean value for small $\st$ in \Figref{2}({\bf c}) agrees well with the optimal fluctuation obtained from the shooting method (dashed line), even though $M=1$, while the shooting method applies strictly only in the limit $M\to\infty$. This suggests that the optimal fluctuation is insensitive to changes of the model when $\st$ is small.
Further evidence for the robustness of the optimal fluctuation is presented in \Figref{2}({\bf d}). The figure shows $\delta\lambda_\text{th}$, the magnitude by which the optimal fluctuation falls below the value $-1/4$ as a function of $\st$, obtained from numerical simulations of the one-dimensional model and from the shooting method. We observe a mild dependence of $\delta\lambda_\text{th}$ on $M$ at finite $\st$, where $M=\infty$ corresponds to the Gaussian model. In the limit $\st\to0$ all threshold values approach $-1/4$ with scaling $\delta\lambda_\text{th}\propto \st^{2/3}$ extracted numerically, albeit with $M$-dependent prefactors. The same $\st^{2/3}$ scaling can also be obtained from a bound derived in B\"atge {\em et al.}~\cite{Bat23}, by using the optimal fluctuation to establish a relation between the duration of exceeding the threshold and $\delta\lambda_\text{th}$, $D$ and $\Delta$ in their Eq.~(8), respectively. We conclude that the optimal fluctuation of fluid-velocity gradients is generally robust against changes of the model.  The rate of caustic formation, by contrast, is sensitive to the tails of the distribution, as shown in \Figref{1}({\bf b}).

The observed robustness of the optimal fluctuation with respect to changes of $M$ justifies using $A^*(t)$ obtained from the shooting method for finite $\st$ as the amplitude for the $d$-dimensional optimal fluctuation in \Eqnref{Aonedim}. With the appropriate parameters $\sigma_S^2 = 2C_S d^{-1}(d+2)^{-1} \sim 1/15$ and $\kappa_S$ obtained from simulations of the non-Gaussian model in three spatial dimensions, $d=3$, we numerically extract the small-$\st$ scaling of $\delta\lambda_{\rm th}$ from the shooting method. The results are given in Table~\figref{scaling}. We observe that $\kappa_S$ depends on $M$, which introduces a weak $M$-dependence of the scaling prefactor of $\delta\lambda_\text{th}$ extracted from the shooting method.
\begin{table}
\renewcommand{\arraystretch}{1.2}
\setlength{\tabcolsep}{8pt}
\setlength\extrarowheight{0pt}
\begin{tabularx}{.48\linewidth}{lcccccc}
	$M$			&	1	&	2	&	5	&	10	&	20	&	50 \\
	\hline
	$\kappa_S$	&	1.33	&	1.57	&	1.72	&	1.81	&	1.82	&	1.82	\\
	$\delta\lambda_\text{th}\st^{-2/3}$&	0.19	&	0.17	&	0.16	&	0.16	&	0.15	&	0.15\\
\end{tabularx}
    \caption{Numerically obtained non-dimensional Lagrangian strain correlation times $\kappa_S$ and scaling prefactors $\delta\lambda_\text{th}\st^{-2/3}$ for different parameters $M$ in the non-Gaussian model.}\figlab{scaling}
\end{table}
\section{Data collapse onto scaling function}\seclab{datacoll}
With the one-dimensional picture \eqnref{Aonedim} of caustic formation developed in the previous section, we now derive the scaling function [black line in \Figref{1}({\bf c})] onto which the caustic formation data collapses for small $\st$. To this end, we again diagonalise $\mbb{A}(t)$ at the time of caustic formation. We then obtain from \Eqnref{3Dmodel} the following expression for the component of the fluid velocity gradient that reaches the threshold $\mbb{A}_\text{th}$:
\algn{
	A(\ve x,t) = \frac1{\sqrt{M}}\sum_{m=1}^M c_m(t) A_m(\ve x)\,,
}
where $A_m = \langle \mbb{A}_m,\mbb{e}\rangle$ is the amplitude of $\mbb{A}_m$ along the basis matrix $\mbb{e}$ in \Eqnref{elementd}. Due to homogeneity and stationarity, the steady-state distribution $P_s(A)$ of $A$, is independent of $\ve x$ and $t$, for any single point in space and time. Furthermore, since $c_m$ and $A_m$ are independent, $A$ follows a large-deviation principle~\cite{touchette2009large} for $M\gg1$,
\algn{\eqnlab{PAs}
	P_s(A) \propto \exp\left[-M \ms{F}\left(\frac{A}{\sqrt{M}\sigma_s \st}\right)\right]\,,
}
with rate function $\ms{F}(a)$. For Gaussian $c_m$ and $A_m$, the function $\ms{F}(a)$ can be obtained explicitly by first computing the generating function of $A$ and applying a Legendre transform~\cite{touchette2009large}. We find
\algn{\eqnlab{ratefunc}
	\ms{F}(a) = \frac12\left\{\sqrt{1+\left(2a\right)^2} - 1	- \ln	\left[\frac{\sqrt{1+\left(2a\right)^2}+1}{2}\right]	\right\}\,.
}
In order for a caustic to form, $A$ in \eqnref{PAs} must reach $-\lambda_\text{th} = -1/4 -\delta\lambda_\text{th}$. With the large-deviation form in \Eqnref{PAs}, this leads us to
\algn{\eqnlab{caustFrel}
	-M^{-1}\ln(\ms{J}\tauk) = \ms{F}\left(\frac{\lambda_\text{th}}{\sqrt{M}\sigma_S\st}\right)\,.
}
Hence, the argument $a$ of the scaling function $\ms{F}$ is given by
\algn{\eqnlab{scalvar}
	a = \frac{1/4 + \delta\lambda_\text{th}}{\sqrt{M}\sigma_S\st}\,.
}
In the limit of $\st\to 0$, Eqs.~\eqnref{caustFrel} and \eqnref{scalvar} simplify to \Eqnref{J2}. When plotted against the parameter $a$, the data collapses onto the scaling function $\ms{F}(a)$ in \Eqnref{ratefunc} as $\st\to 0$, see \Figref{1}({\bf c}).
The inset of \Figref{1}({\bf c}) shows that, for finite values of $\st$, using $\delta\lambda_\text{th}$ from the shooting method with the prefactors in Table~\figref{scaling} improves the collapse, even though the shooting method was derived for $M\to\infty$.

We now show that one obtains \Eqsref{PAG} and \eqnref{PAnG} from the scaling function $\ms{F}$ in different limits. First, for $\st\ll1$ and $\sqrt{M}\st\gg 1$, we have $a\ll1$ in \eqnref{ratefunc}, so that
\algn{\eqnlab{Fasmall}
	M\ms{F}\left(\frac{\lambda_\text{th}}{\sqrt{M}\sigma_S\st}\right)\sim \frac{\lambda_\text{th}^2}{2\sigma_S^2\st^2}\,.
}
Second, for $\st\ll1$ and $\sqrt{M}\st\ll 1$, one has $a\gg 1$, and
\algn{\eqnlab{Falarge}
	M\ms{F}\left(\frac{\lambda_\text{th}}{\sqrt{M}\sigma_S\st}\right)\sim \frac{\sqrt{M}\lambda_\text{th}}{\sigma_S\st}\,.
}
\begin{figure}[t]
\begin{overpic}[width =\textwidth]{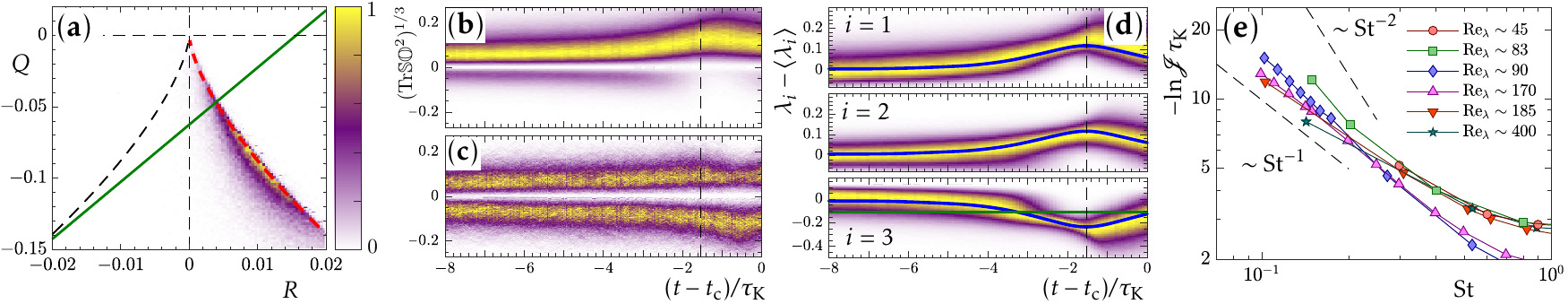}
\end{overpic}
\caption{\label{fig:3} Caustic formation using direct numerical simulations of turbulence to integrate Eqs.~(\ref{eq:st}) and (\ref{eq:Z}).
({\bf a}) Color-coded probability density of $Q$ and $R$ at the onset of caustic formation [$(t_{\rm th}-t_{\rm c})/\tauk=-1.53$].
({\bf b}) Probability density of the real cube root of the invariant $\tr(\ma S\ma O^2)$ before the formation of a caustic at $t=t_{\rm c}$. Dashed line shows the onset time for panel ({\bf a}).
({\bf c}) Same as panel ({\bf b}), but for the statistical model with $M=1$ and $\st=0.1$ [the case in \Figref{1}({\bf a})].
({\bf d}) Probability density of eigenvalues $\lambda_i$ of $\ma S$ minus their average values, $\langle\lambda_i\rangle$, evaluated along particle trajectories independent from the formation of caustics. The solid green line shows the $\st\to 0$ limit of the threshold, $\lambda_\text{th}(\st\to 0)=-1/4$.
Solid blue lines show the optimal fluctuation according to theory for the eigenvalues, $\lambda_3(t)-\langle\lambda_3\rangle\sim \lambda_\text{th}(\st)f_S(t-t_\text{th})$ and $\lambda_1(t)-\langle\lambda_1\rangle=\lambda_2(t)-\langle\lambda_2\rangle\sim -\tfrac{1}{2}\lambda_\text{th}(\st) f_S(t-t_{\rm th})$, with a fitted threshold $\lambda_\text{th}(\st)\approx-0.23$.
Parameters: ${\rm Re}_\lambda=185$, and~$\st=0.3$.
The maximal values of the densities are at each time scaled to unity.
({\bf e}) Rate of caustic formation against Stokes number $\st$. Dashed lines indicate scalings $\st^{-1}$ and $\st^{-2}$.
Data is obtained from Ref.~\cite{falkovich2007sling} (${\rm Re}_\lambda=45,83$), Ref.~\cite{bhatnagar2022rate} (${\rm Re}_\lambda=90,170$), and new results using the DNS underlying Refs.~\cite{bec2006lyapunov} (${\rm Re}_\lambda=185$) and~\cite{bec2010intermittency} (${\rm Re}_\lambda=400$).
}
\end{figure}
These expressions agree with \Eqsref{PAG} and \eqnref{PAnG} at the threshold configuration~\eqnref{Ath}. Hence, the scaling function $\ms{F}$ interpolates between the asymptotic $\st$ scalings derived in \Eqsref{PAG} and \eqnref{PAnG}.
\section{Turbulence}
Finally, we discuss the implication of the non-Gaussian model for DNS of particles in turbulence. To this end, we have analysed data of particle trajectories from DNS~\cite{bec2006lyapunov,bec2010intermittency} and earlier DNS results for the rate of caustic formation~\cite{falkovich2007sling,bhatnagar2022rate}.
Figure~\figref{3}({\bf a}) displays the optimal fluctuation at the time of onset of caustic formation, using the trajectory-data set from Ref.~\cite{bec2006lyapunov}. It shows the trajectory density in the $Q$-$R$-plane at the time $t_\text{th}$ when a caustic is initiated. As in the statistical model data shown in \Figref{1}({\bf a}), the largest probability density lies below the threshold line and close to the right branch of the Vieillefosse line (red dashed line), in qualitative agreement with the theory described above, and with earlier unpublished DNS data~\cite{Bha20}. However, we also see quantitative differences between the shapes of the gradient distributions in \Figref{1}({\bf a}) and \Figref{3}({\bf a}). In particular, the data for turbulent fluid-velocity gradients scatters broadly along and slightly below the Vieillefosse line. For the non-Gaussian model [\Figref{1}({\bf a})], by contrast, the scatter appears more centred on the Vieillefosse line.

The elongated scatter of turbulent fluid-velocity gradients is likely due the fact that the non-linear terms in the Navier-Stokes equations self amplify fluid-velocity gradients along this branch~\cite{vieillefosse1982local,vieillefosse1984internal}. This generates large gradient excursions and thus leads to anomalously large fluctuations along the right branch of the Vieillefosse line.
That the turbulent gradients scatter below the Vieillefosse line, instead of centering on the line as in the non-Gaussian model, can be understood by decomposing $Q=-{\rm tr}\ma S^2/2+{\rm tr}\ma O\ma O^{\rm T}/2$ and $R=-{\rm tr}\ma S^3/3-{\rm tr}\ma S\ma O^2$. If $\ma O=0$, no pair of ($R$,$Q$) may lie above the Vieillefosse line by its definition.
In this case, the distribution is concentrated just below the Vieillefosse line, for both the Gaussian model and the DNS (not shown).
The effect of non-zero $\ma O$ is to first shift $Q$ upwards, because ${\rm tr}\ma O\ma O^{\rm T}\ge 0$.
Second, a negative value of ${\rm tr}\ma S\ma O^2$ shifts $R$ to the right, and vice versa.
This explains why the DNS data lies below the Vieillefosse line in the $Q-R$ plane. The reason is that ${\rm tr}\ma S\ma O^2$ is unlikely to take negative values. This is apparent from Figure~\figref{3}({\bf b}), which shows that the probability to have negative ${\rm tr}\ma S\ma O^2$ is very small at the onset of caustic formation (the onset time is indicated by the vertical dashed line).
This peculiar coupling between strain and vorticity as the strain grows large is specific to turbulence.
In the statistical model, the distribution of ${\rm tr}\ma S\ma O^2$ is almost symmetric, see \Figref{3}({\bf c}), explaining why the data in \Figref{1}({\bf a}) is scattered both above and below the Vieillefosse line.

The strong interrelation between strain and vorticity in turbulence discussed above also results in vorticity giving a finite, correlated contribution to the optimal fluctuation of $Q$ (not shown)~\cite{baetge2023private}. This is in contrast to the Gaussian and non-Gaussian models considered here, where the optimal vorticity contribution vanishes~\cite{meibohm2021paths,meibohm2023caustics}.
Despite these differences, the shape of the optimal fluctuation in the DNS is still quite well approximated by the correlation function, as shown in \Figref{3}({\bf d}), where the trajectory density of the eigenvalues of the turbulent strain $\mbb{S}$, are depicted prior to caustic formation.
The optimal fluctuation theory described above assumes that the mean values $\langle\lambda_i\rangle$ vanish.
This is consistent, because they are of higher order than $\delta\lambda\sim\st ^{2/3}$ for small $\st$.
Since the Stokes number in \Figref{3}({\bf d}) is not very small, we subtracted the mean values from $\lambda_i$. The solid blue lines show the theoretical predictions for the optimal fluctuations, given by the correlation function $f_S$ of $\mbb{S}$ normalised to a fitted threshold value. We observe qualitative agreement.

Finally, \Figref{3}({\bf e}) shows the $\st$-dependence of rate of caustic formation $\ms{J}\tauk$ for particles in turbulence from Refs.~\cite{bhatnagar2022rate,falkovich2007sling}, using DNS with different Reynolds numbers $\text{Re}_\lambda$. In the literature, different authors have fitted different exponents, ranging between minus one and minus two, to the $\st$-dependence of $-\ln \mathscr{J}\tauk$~\cite{bhatnagar2022rate,falkovich2007sling}, motivated by
results for the rate of caustic formation in Gaussian statistical models \cite{gustavsson2016statistical}.
The data in \Figref{3}({\bf e}) appears to be consistent with inverse linear scaling $\st^{-1}$ for some of the cases, but the scaling exponent depends strongly on $\text{Re}_\lambda$. This is consistent with the observations in Ref.~\cite{Bat23} and supported by the analysis of the non-Gaussian statistical model: the small-$\st$ scaling of $-\ln \mathscr{J}\tauk$ is \text{strongly affected by} the extreme-value statistics of fluid-velocity gradients in turbulence, not captured by  Gaussian models.
\section{Conclusions}
We formulated a non-Gaussian statistical model for turbulent aerosols, with non-Gaussian tails of the distribution
of  fluid-velocity gradients. We analysed how caustics form in this model at small Stokes numbers, in the persistent limit. Using large-deviation theory we computed the rate of caustic formation for the non-Gaussian model and demonstrated that caustic formation is sensitive to the tails of the distribution of fluid-velocity gradients, because they must overcome a threshold for a caustic to form: The smaller the Stokes number, the larger the fluid-velocity gradients must be to initiate a caustic. For $\st\lessapprox 0.1$ the required fluid-velocity gradients lie far in the tails of the distribution.

While the rate of caustic formation depends sensitively on the tails of the fluid-velocity gradient distribution seen by the particles, the form of the optimal fluctuation is robust against changes of the parameters in the non-Gaussian model. In particular, the optimal fluctuation of the non-Gaussian model has small vorticity, just like the Gaussian one, and it follows the right branch of the Vieillefosse line at small Stokes numbers.

What are the implications of these results for caustics in three-dimensional homogeneous isotropic turbulent flow? First, our analysis of the non-Gaussian model shows that
the rate of caustic formation is sensitive to the tails of the fluid-velocity gradient distribution. As a consequence, the rate of caustic formation is not expected to have a simple scaling form such as $\st^{-1}$ or $\st^{-2}$. This explains why numerical studies of the rate of caustic formation using direct numerical simulation of turbulence at different Reynolds numbers yielded inconclusive results~\cite{bhatnagar2022rate,falkovich2007sling}, simply because the $\st$-dependence of $-\ln  (\mathscr{J}{{\tauk}})$ is not a power law. These conclusions are consistent with the results of Ref.~\cite{Bat23}, who analysed the  Reynolds-number dependence of the rate of caustic formation using DNS of turbulence.
Second, the optimal fluctuation of fluid-velocity gradients leading to caustic formation in turbulence is similar to the predictions of our non-Gaussian model.
In particular, the fluid-velocity gradients that initiate caustics in turbulence lie close to the right branch of the Vieillefosse line. This robustness explains why particle collisions
in turbulence tend to occur close to the Vieillefosse line~\cite{Lee23}. Third, the vorticity contribution to the optimal fluctuation in turbulence, however, is not simply scattered around zero but performs a small, yet correlated excursion. This suggests an interdependence of vorticity and strain in the tails of the joint distribution of turbulent strain and vorticity~\cite{Bat23}. We remark that one can produce an optimal fluctuation with non-zero vorticity by adding third- and fourth-order terms to the (quadratic) Gaussian action in \Eqnref{PAG}. Whether or not the finite optimal vorticity in turbulence has a similar origin is an open question.
\section{Acknowledgments}
The research of LS and BM was supported in part by VR grant 2021-4452. KG was supported by VR Grant No. 2018-03974. JM was funded by a Feodor-Lynen Fellowship of the Alexander von Humboldt-Foundation. The statistical-model simulations were conducted using the resources of HPC2N provided by the Swedish National Infrastructure for Computing (SNIC), partially funded by the Swedish Research Council through grant agreement no. 2018-05973. We thank Luca Biferale for granting access to the DNS data needed to generate the results in Fig. 3.

%

\begin{thebibliography}{10}

\bibitem{bec2003fractal}
J.~Bec.
\newblock Fractal clustering of inertial particles in random flows.
\newblock {\em Phys. Fluids}, 15:L81--L84, 2003.

\bibitem{crisanti1992lagrangian}
A.~Crisanti, M.~Falcioni, A.~Provenzale, P.~Tanga, and A.~Vulpiani.
\newblock Dynamics of passively advected impurities in simple two-dimensional
  flow models.
\newblock {\em Phys. Fluids}, 4:1805--1820, 1992.

\bibitem{falkovich2002acceleration}
G.~Falkovich, I.~Fouxon, and Stepanov M.
\newblock Acceleration of rain initiation by cloud turbulence.
\newblock {\em Nature}, 419:151--154, 2002.

\bibitem{wilkinson2005caustics}
M.~Wilkinson and B.~Mehlig.
\newblock Caustics in turbulent aerosols.
\newblock {\em Europhys. Lett.}, 71:186--192, 2005.

\bibitem{gustavsson2016statistical}
K.~Gustavsson and B.~Mehlig.
\newblock Statistical models for spatial patterns of heavy particles in
  turbulence.
\newblock {\em Adv. Phys.}, 65:1, 2016.

\bibitem{bec2024statistical}
J.~Bec, K.~Gustavsson, and B.~Mehlig.
\newblock Statistical models for the dynamics of heavy particles in turbulence.
\newblock {\em Annual Reviews of fluid mechanics}, 56:189--213, 2024.

\bibitem{bec2010intermittency}
J.~Bec, L.~Biferale, M.~Cencini, A.~Lanotte, and F.~Toschi.
\newblock Intermittency in the velocity distribution of heavy particles in
  turbulence.
\newblock {\em J. Fluid Mech.}, 646:527--536, 2010.

\bibitem{salazar2012inertial}
Juan P. L.~C. Salazar and Lance~R. Collins.
\newblock Inertial particle relative velocity statistics in homogeneous
  isotropic turbulence.
\newblock {\em J. Fluid Mech.}, 696:45--66, 2012.

\bibitem{gustavsson2014relative}
K.~Gustavsson and B.~Mehlig.
\newblock {Relative velocities of inertial particles in turbulent aerosols}.
\newblock {\em J. Turbul.}, 15:34--69, 2014.

\bibitem{falkovich2007sling}
G.~Falkovich and A.~Pumir.
\newblock Sling effect in collisions of water droplets in turbulent clouds.
\newblock {\em J. Atmos. Sci}, 64:4497--4505, 2007.

\bibitem{bewley2013observation}
Gregory Bewley, Ewe~Wei Saw, and Eberhard Bodenschatz.
\newblock Observation of the sling effect.
\newblock {\em New Journal of Physics}, 15:083051, 08 2013.

\bibitem{fevrier2005partitioning}
P.~Fevrier, O.~Simonin, and K.~D.~Squires.
\newblock Partitioning of particle velocities in gas-solid turbulent flows into a continuous field and a spatially uncorrelated random distribution: theoretical formalism and numerical study.
\newblock {\em J.~Fluid~Mech.}, 533:1, 2005.

\bibitem{ducasse2009inertial}
L.~Ducasse, and A.~Pumir.
\newblock Inertial particle collisions in turbulent synthetic flows: Quantifying the sling effect.
\newblock {\em Phys.~Rev.~E}, 80:066312, 2009.

\bibitem{ijzermans2010segregation}
R.~H.~A.~Ijzermans, E.~Meneguz, and M.~W.~Reeks.
\newblock Segregation of particles in incompressible random flows: singularities, intermittency and random uncorrelated motion.
\newblock {\em J.~Fluid~Mech.}, 653:99, 2010.

\bibitem{papoutsakis2018modelling}
A~Papoutsakis, O.~D.~Rybdylova, T.~S.~Zaripov, L.~Danaila, A.~N.~Osiptsov, and S.~S.~Sazhin.
\newblock Modelling of the evolution of a droplet cloud in a turbulent flow.
\newblock {\em Int.~J.~Multiph.~Flow.}, 104:233, 2018.

\bibitem{meibohm2021paths}
Jan Meibohm, Vikash Pandey, Akshay Bhatnagar, Kristian Gustavsson, Dhrubaditya
  Mitra, Prasad Perlekar, and Bernhard Mehlig.
\newblock Paths to caustic formation in turbulent aerosols.
\newblock {\em Phys. Rev. Fluids}, 6:L062302, 2021.

\bibitem{meibohm2023caustics}
Jan Meibohm, Kristian Gustavsson, and Bernhard Mehlig.
\newblock Caustics in turbulent aerosols form along the {V}ieillefosse line at
  weak particle inertia.
\newblock {\em Phys. Rev. Fluids}, 8:024305, 2023.

\bibitem{derevyanko2007lagrangian}
SA~Derevyanko, Gregory Falkovich, K~Turitsyn, and S~Turitsyn.
\newblock Lagrangian and eulerian descriptions of inertial particles in random
  flows.
\newblock {\em Journal of Turbulence}, 8:N16, 2007.

\bibitem{meibohm2017relative}
J.~Meibohm, L.~Pistone, K.~Gustavsson, and B.~Mehlig.
\newblock Relative velocities in bidisperse turbulent suspensions.
\newblock {\em Phys. Rev. E}, 96:061102, 2017.

\bibitem{meibohm2019heavy}
J~Meibohm and B~Mehlig.
\newblock Heavy particles in a persistent random flow with traps.
\newblock {\em Physical Review E}, 100(2):023102, 2019.

\bibitem{gustavsson2013distribution}
K.~Gustavsson and B.~Mehlig.
\newblock Distribution of velocity gradients and rate of caustic formation in
  turbulent aerosols at finite {K}ubo numbers.
\newblock {\em Phys. Rev. E}, 87:023016, 2013.

\bibitem{bec2014gravity}
J.~Bec, H.~Homann, and S.~Sankar~Ray.
\newblock Gravity-driven enhancement of heavy particle clustering in turbulent
  flow.
\newblock {\em Phys. Rev. Lett.}, 112:184501, 2014.

\bibitem{gustavsson2014clustering}
K.~Gustavsson, S.~Vajedi, and B.~Mehlig.
\newblock Clustering of particles falling in a turbulent flow.
\newblock {\em Phys. Rev. Lett.}, 112:214501, 2014.

\bibitem{papoutsakis2020a}
A.~Papoutsakis, and M.~Gavaises
\newblock A model for the investigation of the second-order structure of caustic formations in dispersed flows.
\newblock {\em J. Fluid Mech.}, 892:A4, 2020.

\bibitem{Bat23}
Tobias B{\"{a}}tge, Itzhak Fouxon, and Michael Wilczek.
\newblock {Quantitative prediction of sling events in turbulence at high
  Reynolds numbers}.
\newblock {\em Phys. Rev. Lett.}, 131:054001, 2023.

\bibitem{falkovich2001particles}
G.~Falkovich, K.~Gaw\c{e}dzki, and M.~Vergassola.
\newblock Particles and fields in fluid turbulence.
\newblock {\em Rev. Mod. Phys.}, 73:913--975, 2001.

\bibitem{maxey1987gravitational}
M.~R. Maxey.
\newblock The gravitational settling of aerosol particles in homogeneous
  turbulence and random flow fields.
\newblock {\em J. Fluid Mech.}, 174:441--465, 1987.

\bibitem{girimaji1990diffusion}
SS~Girimaji and SB~Pope.
\newblock A diffusion model for velocity gradients in turbulence.
\newblock {\em Physics of Fluids A: Fluid Dynamics}, 2(2):242--256, 1990.

\bibitem{brunk1997hydrodynamic}
Brett~K Brunk, Donald~L Koch, and Leonard~W Lion.
\newblock Hydrodynamic pair diffusion in isotropic random velocity fields with
  application to turbulent coagulation.
\newblock {\em Physics of Fluids}, 9(9):2670--2691, 1997.

\bibitem{zaichik2003pair}
L.~I. Zaichik and V.~M. Alipchenkov.
\newblock Pair dispersion and preferential concentration of particles in
  isotropic turbulence.
\newblock {\em Phys. Fluids}, 15:1776, 2003.

\bibitem{vincenzi2007stretching}
Dario Vincenzi, Shi Jin, Eberhard Bodenschatz, and Lance~R Collins.
\newblock Stretching of polymers in isotropic turbulence: a statistical
  closure.
\newblock {\em Physical Review Letters}, 98(2):024503, 2007.

\bibitem{squires1991preferential}
K.~D. Squires and J.~K. Eaton.
\newblock Preferential concentration of particles by turbulence.
\newblock {\em Phys. Fluids A}, 3:1169--1178, 1991.

\bibitem{eaton1994preferential}
J.K. Eaton and J.R. Fessler.
\newblock Preferential concentration of particles by turbulence.
\newblock {\em Int. J. Multiphase Flow}, 20:169--209, 1994.

\bibitem{touchette2009large}
Hugo Touchette.
\newblock The large deviation approach to statistical mechanics.
\newblock {\em Physics Reports}, 478(1-3):1--69, 2009.

\bibitem{perrin2014preferred}
Vincent~E. Perrin and Harm J.~J. Jonker.
\newblock Preferred location of droplet collisions in turbulent flows.
\newblock {\em Phys. Rev. E}, 89:033005, 2014.

\bibitem{picardo2019flow}
Jason~R Picardo, Lokahith Agasthya, Rama Govindarajan, and Samriddhi~Sankar
  Ray.
\newblock Flow structures govern particle collisions in turbulence.
\newblock {\em Physical Review Fluids}, 4(3):032601, 2019.

\bibitem{Lee23}
S.~Lee and C.~Lee.
\newblock Identification of a particle collision as a finite-time blowup in
  turbulence.
\newblock {\em Sci. Rep.}, 13:181, 2023.

\bibitem{bhatnagar2022rate}
Akshay Bhatnagar, Vikash Pandey, Prasad Perlekar, and Dhrubaditya Mitra.
\newblock Rate of formation of caustics in heavy particles advected by
  turbulence.
\newblock {\em Philosophical Transactions of the Royal Society A: Mathematical,
  Physical and Engineering Sciences}, 380:20210086, 2022.

\bibitem{bec2006lyapunov}
J.~Bec, L.~Biferale, G.~Boffetta, M.~Cencini, S.~Musacchio, and F.~Toschi.
\newblock Lyapunov exponents of heavy particles in turbulence.
\newblock {\em Phys. Fluids}, 18:091702, 2006.

\bibitem{Bha20}
Akshay Bhatnagar.
\newblock unpublished, 2020.

\bibitem{vieillefosse1982local}
P~Vieillefosse.
\newblock Local interaction between vorticity and shear in a perfect
  incompressible fluid.
\newblock {\em Journal de Physique}, 43(6):837--842, 1982.

\bibitem{vieillefosse1984internal}
P~Vieillefosse.
\newblock Internal motion of a small element of fluid in an inviscid flow.
\newblock {\em Physica A: Statistical Mechanics and its Applications},
  125(1):150--162, 1984.

\bibitem{baetge2023private}
T.~B{\"a}tge.
\newblock private communication.

\end{thebibliography}

\appendix

\section{Derivation of Eq.~(\ref{eq:J2})}
\label{app:A}
Here we show that the rate of caustic formation for  the non-Gaussian model~\eqnref{3Dmodel} is of the form~(\ref{eq:J2}) at small $\st$. To this end, we express $\mbb{A}$ in terms of the uncorrelated components
\algn{\eqnlab{matexp}
	A^i(t) = \langle\mbb{e}^i, \mbb{A}(t)\rangle\,,
}
where $\mbb{e}_i$, $i=1,\ldots,N_d$ furnish a basis of traceless matrices, such that $\langle\mbb{e}^i,\mbb{e}^j\rangle=\delta_{ij}$, with the inner product defined in \Secref{1D} and $\mbb{e}^{N_s}\equiv \mbb{e}$ given in \Eqnref{elementd}. These $N_d=d^2-1$ basis elements can be split into a set of $N_s = (d-1)(d+2)/2$ symmetric and $N_O = d(d-1)/2$ antisymmetric traceless matrices, where $d$ is the spatial dimension. From the non-Gaussian model~\eqnref{3Dmodel}, we then obtain
\algn{
	A^i(\ve x,t) = \frac1{\sqrt{M}}\sum_{m=1}^M c_m(t) A^i_m(\ve x)\,.
}
For different $m$, the $A^i_m(\ve x) = \langle\mbb{e}^i, \mbb{A}_m\rangle$ are uncorrelated Gaussian random variables with zero mean, and thus independent. We define $\ve A = (A^1,\ldots,A^{N_d})^{\sf T} = (\ve S,\ve O)^{\sf T}$, where $\ve S = (S^1,\ldots,S^{N_S})^{\sf T}$ and $\ve O = (O^1,\ldots,O^{N_O})^{\sf T}$. The vectors $\ve S$ and $\ve O$ correspond to the strain $\mbb{S}$ and vorticity $\mbb{O}$ parts of $\mbb{A}$, respectively. 
We also introduce $\boldsymbol{A}_m = (A_m^1,\ldots,A_m^{N_d})^{\sf T}=(\boldsymbol{S}_m,\boldsymbol{O}_m)^{\sf T}$, where $\boldsymbol{S}_m = (S_m^1,\ldots,S_m^{N_S})^{\sf T}$ and $\boldsymbol{O}_m = (O_m^1,\ldots,O_m^{N_O})^{\sf T}$.
The latter have the covariances
\algn{
	\langle S^i_m(\ve x) S^j_n(\ve x)\rangle = \delta_{nm}\delta^{ij}\st^2\sigma_S^2\,,\qquad  \langle O^i_m(\ve x) O^j_n(\ve x)\rangle = \delta_{nm}\delta^{ij}\st^2\sigma_O^2\,,
}
and $\langle S^i_m(\ve x) O^j_n(\ve x)\rangle = 0$, where $\sigma_S^2$ is given in \Eqnref{sigS} and
\algn{
	\sigma_O^2 = \frac{2C_O(\st)}{d^2}\,.
}
For the following calculation, it is convenient to introduce the variables $\tilde{\ve A} = (\boldsymbol{S},\boldsymbol{O}\sigma_S/\sigma_O)^{\sf T}$ and $\ve{\tilde{A}}_m = (\ve S_m,\ve O_m \sigma_S/\sigma_O)$, so that all $\tilde A^i$ and $\tilde{A}^i_m$ with $i=1,\ldots,N_d$ have the same variance $\sigma_S$. We now write the time and space independent steady-state distribution $P_s(\mbb{A})$ in terms of $\ve S$ and $\ve O$. To this end, we first compute $P_s(\tilde{\ve A})$ and transform the result to obtain $P_s(\mbb{A})$ for $\mbb{A}\gg\st$. We have
\algn{\eqnlab{psso}
	P_s(\tilde{\ve A}) = \left\langle \prod_{i=1}^{N_S}\delta\left(\tilde{A}^i - \frac{\ve c \cdot \ve A^i}{\sqrt{M}}\right)\prod_{i=1}^{N_O}\delta\left(\tilde{A}^{N_S+i} - \frac{\ve c \cdot \ve A^{N_S+i}\sigma_S/\sigma_O}{\sqrt{M}}\right)\right\rangle\,,
}
where $\ve A^i = (A^i_1,\ldots,A^i_M)^{\sf T}$ and $\ve c = (c_1,\ldots,c_M)^{\sf T}$. The random variables $A^i_m$ and $c_m(t)$ are independent and Gaussian. After the change of variables $\ve c\to\ve c/\sqrt{\sigma_S}$, $\ve A^{i}\to \ve A^{i}\sqrt{\sigma_S}$, $\ve A^{N_S+i}\to \ve A^{N_S+i}\sigma_O/\sqrt{\sigma_S}$, the average in \Eqnref{psso} is expressed as the multi-dimensional Gaussian integral:
\begin{widetext}
\algn{\eqnlab{psso2}
	P_s(\tilde{\ve A}) = N_{\ve A}N_{\ve c}\int_{\mbb{R}^{M\times N_d}}\!\!\prod_{i=1}^{N_d}\ed\ve A^i\!\!\int_{\mbb{R}^M}\!\!\ed\ve c\prod_{i=1}^{N_d}\delta\left(\tilde{A}^i - \frac{\ve c\cdot \ve A^i}{\sqrt{M}}\right) \ee^{-\frac1{2\sigma_S\st^2}\left(|\ve c|^2+\sum_{i=1}^{N_d}|\ve A^i|^2\right)}\,,
}
\end{widetext}
where $N_{\ve A}$ and $N_{\ve c}$ are appropriate normalisation factors. For $\st\ll1$ the integral can be evaluated by a saddle-point approximation, subject to the constraints in the $\delta$-functions. We thus minimise the Lagrange function
\algn{\eqnlab{Lagrangian}
	\ms{L} = \frac{|\ve c|^2}{2}&+\sum_{i=1}^{N_d}\frac{|\ve A^i|^2}{2} - \sum_{i=1}^{N_d}\beta_i\left(\tilde{A}^i - \ve c\cdot \ve A^i/\sqrt{M}\right)\,,
}
where $\beta_i$ are Lagrange multipliers that enforce the constraints. Minimisation
\algn{\eqnlab{speqn}
	\frac{\partial \ms{L}}{\partial \ve A^{i}}\bigg|_{\ve A^*,\ve c^*} = 0\,,\qquad \frac{\partial \ms{L}}{\partial \ve c}\bigg|_{\ve A^*,\ve c^*} = 0\,,
}
leads to the saddle-point equations
\algn{\eqnlab{speqn2}
	\ve A^{*i} + \frac{\beta_i\ve c^{*i}}{\sqrt{M}} = 0\,, \qquad \ve c^*  + \sum_{j=1}^{N_d}\frac{\beta_j\ve A^{*j}}{\sqrt{M}}=0\,.
}
We take the inner product of $\ve A^{*i}$ and the second equation in \Eqnref{speqn2}. Using the constraints on the inner product gives
\algn{
	\tilde{A}^i + \frac1M\sum_{j=1}^{N_d} \beta_j \ve A^{*i}\cdot\ve A^{*j}  = 0\,,
}
for $i=1,\ldots,N_d$. Taking the inner product of the first equation in \Eqnref{speqn2} with $\ve A^{*i}$ and $\ve A^{*j}$, $j\neq i$, and substituting the result into the previous equation, gives the relations
\algn{
	|\ve \beta|^2 = M\,, \qquad \beta_i \tilde{A}^j = \beta_j \tilde{A}^i\,,
}
where $\ve \beta = (\beta_1,\ldots,\beta_{N_d})^{\sf T}$. Solving these equations for $\ve \beta$ gives
\algn{
	\beta_i = -\frac{\sqrt{M}\tilde{A}^i}{\sqrt{\sum_{j=1}^{N_d}(\tilde{A}^j)^2}}\,.
}
Further, substituting \Eqsref{speqn2} into \Eqnref{Lagrangian}, we obtain
\algn{
	\ms{L} = -\sum_{i=1}^{N_d}\beta_i \tilde{A}^i = \sqrt{M|\ve\tilde{A}|^2}\,.
}
Transforming back, $\ve\tilde{A}\to(\ve S,\ve O)$, we find that the joint probability density of $\ve S$ and $\ve O$ takes the form
\algn{
	-\ln P(\ve S,\ve O)\sim\st^{-1}\sqrt{M\left(|\ve S|^2/\sigma_S^2+|\ve O|^2/\sigma_O^2\right)}\,.
}
Hence, up to the prefactor $\sqrt{M}$ and a factor of $2$, the asymptotic action of the components $A^i$ in the non-Gaussian model is equal to the square-root of the Gaussian action. This shows that $\mathscr{H}(x) = \sqrt{2M x}$ for $|A_{ij}|\gg\sqrt{M}$ as stated in the main text.

\section{Optimal fluctuation}
\label{app:B}
Here we analyse the shape of the optimal fluctuation $\mbb{A}^*(t)$ of fluid-velocity gradients in the non-Gaussian model for $\st\to0$ in terms of the amplitude $A^*(t)$ given in \Eqnref{Aonedim}. For Gaussian processes, an optimal fluctuation $A^*(t)$ that reaches a given threshold $|\lambda_{\rm th}|\gg \sigma_S$ at time $t=t_\text{th}$ is given by the correlation function, $A^*(t) = \lambda_\text{th} f_S(t-t_\text{th})$, normalised to the threshold value $\lambda_\text{th}$~\cite{meibohm2021paths,meibohm2023caustics}. We now show that the optimal fluctuation has the same form for the non-Gaussian model when $\st\to0$.

Since we are dealing with a one-dimensional process, it is enough to consider single component $A(t)\equiv A^{N_s}(t)$ along $\mbb{e}^{N_s}\equiv\mbb{e}$. For the probability $P_s(A=-\lambda_\text{th})$, this is equivalent to considering the Gaussian integral in \Eqnref{psso2} with $N_d=1$. For $|\lambda_\text{th}|\gg \sqrt{M}$, the integral is dominated by the saddle point $(\ve A^*,\ve c^*)$, where $\ve A^* = ({A_1^*}^{N_S},\ldots, {A_M^*}^{N_S})^{\sf T}$ and $\ve c^* = ({c_1^*}^{N_S},\ldots, {c_M^*}^{N_S})^{\sf T}$. Since the process $\ve c(t)$ is Gaussian, $\ve c^*(t)$ reaches $\ve c^*$ at time $t$ by an optimal fluctuation $\ve c^*(t)$ of the form~\cite{meibohm2021paths,meibohm2023caustics}
\algn{
	\ve c^*(t) = \ve c^* f_S(t-t_\text{th})\,.
}
The saddle point $(\ve A^*,\ve c^*)$ is determined by \Eqsref{speqn2}. The optimal fluctuation $A^*(t)$ is then obtained by taking the inner product with $\ve A^*$ and dividing by $\sqrt{M}$:
\algn{
	A^*(t) = \frac{1}{\sqrt{M}}\ve A^*\cdot\ve c^* f_S(t-t_\text{th}) = -\lambda_\text{th}f_S(t-t_\text{th})\,.
}
This shows that in the limit $\st\to0$, the optimal fluctuation $A^*(t)$ for the non-Gaussian process $A(t)$ that reaches $-\lambda_\text{th}$ at time $t$ has, just as in the Gaussian case, the shape of the correlation function of the constituting processes $\ve c(t)$, normalised to the threshold value $-\lambda_\text{th}$, as stated in the main text.

\end{document}